\documentclass[modern]{aastex61}

\usepackage{newtxtext,newtxmath}

\newcommand{\be}{\begin{equation}}
\newcommand{\ee}{\end{equation}}
\newcommand{\ba}{\begin{eqnarray}}
\newcommand{\ea}{\end{eqnarray}}
\newcommand{\lya}{Ly$\alpha$}
\newcommand{\hMpc}{\ h^{-1}\text{Mpc}}

\newcommand{\ihMpc}{\ h\text{Mpc}^{-1}}
\newcommand{\kpar}{k_\parallel}

\newcommand{\HIDE}[1]{}

\newcommand{\LAHIDE}[1]{}

\newcommand{\AFRHIDE}[1]{}

\newcommand{\APHIDE}[1]{}

\newcommand{\FVHIDE}[1]{}

\newcommand{\KRHIDE}[1]{}

\newcommand{\SGHIDE}[1]{}
\submitjournal{ApJ}

\shorttitle{Suppressed Variance in \lya\ Forest Simulations}
\shortauthors{Anderson et al.}

\begin{document}

\title{Cosmological Hydrodynamic Simulations with Suppressed Variance in the Lyman-$\alpha$ Forest Power Spectrum}

\correspondingauthor{Lauren Anderson}
\email{landerson@flatironinstitute.org}

\author[0000-0001-5725-9329]{Lauren Anderson}
\affil{Flatiron Institute 162 5th Ave New York, NY 10010, USA}

\author{Andrew Pontzen}
\affil{University College London, Gower St, Kings Cross, London WC1E 6BT}

\author{Andreu Font-Ribera}
\affil{University College London, Gower St, Kings Cross, London WC1E 6BT}

\author{Francisco Villaescusa-Navarro}
\affil{Flatiron Institute 162 5th Ave New York, NY 10010, USA}

\author{Keir K. Rogers}
\affil{Oskar Klein Centre for Cosmoparticle Physics, Stockholm University, Stockholm SE-106 91, Sweden}

\author{Shy Genel}
\affil{Flatiron Institute 162 5th Ave New York, NY 10010, USA}

\begin{abstract}

We test a method to reduce unwanted sample variance when predicting
Lyman-$\alpha$ (\lya) forest power spectra from cosmological hydrodynamical simulations.
Sample variance arises due to sparse sampling of modes on large scales and propagates to small scales through non-linear gravitational evolution. To tackle this, we generate initial conditions in which the density perturbation amplitudes are {\it fixed} to the ensemble average power spectrum -- and are generated in {\it pairs} with exactly opposite phases.  We run $50$ such simulations ($25$ pairs) and compare their performance against $50$ standard simulations by measuring the \lya\ 1D and 3D power spectra at redshifts $z=2$, 3, and 4. Both ensembles use periodic boxes of $40\hMpc$ containing $512^3$ particles each of dark matter and gas. 
As a typical example of improvement, for wavenumbers $k=0.25\ihMpc$ at $z=3$, we find estimates of the 1D and 3D power spectra converge $34$ and $12$ times faster in a paired-fixed ensemble compared with a standard ensemble.
We conclude that, by reducing the computational time required to achieve fixed accuracy on predicted power spectra, the method frees up resources for exploration of varying thermal and cosmological parameters -- ultimately allowing the improved precision and accuracy of statistical inference.

\end{abstract}


\section{Introduction} \label{sec:intro}

Precision cosmology with the Lyman-$\alpha$ forest is an area of increasing interest, both for its
ability to constrain the evolution of the Universe \citep{McDonald2007,Busca2013,Slosar2013,
Font-Ribera2014a,Bautista2017,duMasdesBourboux2017}
and for its unique insight into the
small-scale power spectrum that is a potential diagnostic of warm dark matter, neutrinos, and other particle physics phenomena
\citep{Seljak2005,Seljak2006b,Seljak2006a,Viel2006b,Viel2013,Palanque-Delabrouille2015,Irsic2017}.
For the latter studies it is essential to explore the effect of both particle physics and
astrophysical parameters on the measured power spectrum, since there are important degeneracies with the
thermal and ionization history of the intergalactic medium (IGM) that need to be considered
\citep{McDonald2005b,McQuinn2016,Walther2018b}.

Ideally one runs a series of cosmological simulations to forward model the parameters into mock observables that can be compared with data \citep{McDonald2005a,Viel2006a,Borde2014}.
However, the mock observations suffer from intrinsic sample variance due to the finite simulation box size; sparse sampling of
random phases and amplitudes of large-scale modes in the initial conditions propagate to an uncertainty in the final mock power
spectrum. This uncertainty can affect even small-scale modes through non-linear gravitational coupling. In practice, model uncertainty is the limiting factor on large scales where it exceeds the observational uncertainty \citep{Palanque-Delabrouille2015}
Suppressing the uncertainties can be achieved either by expanding the box size, or by averaging many simulated realizations of the same volume size; either way
a more precise mean power spectrum (or other observable) can be achieved. But each simulation is expensive, especially as the
box size is increased. To extract the best possible constraints, one has to tension this need for large sample sizes or volumes for
a single point in  parameter space against the requirement to adequately sample the parameter space itself.

This paper examines an approach to lessen the tension, by initializing simulations with carefully chosen amplitudes and
phases for each mode, thus reducing the total volume or number of simulations required to achieve a given statistical accuracy on the
mean mock power spectrum. This approach will free up computing time per parameter instantiation to increase the density of samples in parameter space. Methods, such as emulators (Bird et al. in prep.; Rogers et al. in prep.), which interpolate between hydrodynamic simulations predictions, can then be more accurate. Interpolation is  necessary because only a limited number of simulations can feasibly be run.

We minimize the number of simulation realizations by running {\it paired, fixed} simulations \citep{Pontzen2016, Angulo2016}.
The difference between the standard simulations and the paired-fixed simulations is in the input density field.
For a Gaussian random field, the amplitude of each mode is drawn from a Rayleigh distribution with zero mean and variance equal to the power spectrum at that scale, and the phase drawn with uniform probability between 0 and $2 \pi$.
Instead, for paired-fixed simulations, we fix the amplitudes of the Fourier modes to the square root of the power spectrum, and invert the phase of one simulation in a pair relative to the other.
The fixing reduces the variance of the power spectrum on the largest scales, which remain in or near the linear regime, while the pairing cancels leading-order errors due to non-linear evolution of chance correlations between large-scale modes.

On its own pairing is a benign manipulation in the sense that both realisations in the pair are legitimate, equally-likely draws from a Gaussian random field \citep{Pontzen2016}. Fixing, on the other hand, generates an ensemble that is quantitatively different from Gaussian; \cite{Angulo2016} argued that, to any order in perturbation theory, the differences affect only the variance on the power spectrum and cannot propagate to other observables such as the power spectrum or  observable measures of non-Gaussianity. Combined, pairing and fixing thus formally removes the leading order and next-to-leading order uncertainties that generate scatter in predicted mock observations from simulations.  In \cite{Villaescusa2018}, we thoroughly explored the statistics of paired-fixed simulations for both dark matter evolution and galaxy formation, finding that there are no statistically significant biases in the approach, and in many settings it does yield significantly enhanced statistical accuracy.

In this paper, we apply these techniques for the first time to the study of the power spectrum of fluctuations in the Lyman-$\alpha$ (\lya) forest, as measured from hydrodynamical simulations. The \lya\ forest is the collection of absorption features present in distant quasar spectra due to intervening neutral hydrogen absorbing \lya\ photons as the quasar light redshifts. Physically, the forest comprises a relatively smooth, low density environment of hydrogen which traces the underlying matter density of the Universe well in the redshift range $2<z<5$ \citep{Croft1998, 1996ApJ...457L..51H}. At higher redshifts $z>5$ and especially beyond reionization the neutral hydrogen is too optically thick \citep{1965ApJ...142.1633G}; at lower redshift $z<2$ the forest cannot be observed with optical instruments. Although use of the \lya\ forest as a tracer of the matter density is limited to a finite redshift interval, it is a unique probe of a wide range of otherwise inaccessible redshifts and  scales. It has the unique advantage of tracing the matter density at small scales, down to 10s of kpc, in regions where the evolution is still close to linear.

This paper is structured as follows.
In Section \ref{sec:sims} we describe the simulations used in this analysis,
before presenting results in Section \ref{sec:results}.
We conclude in Section \ref{sec:conc} with a brief summary and discussion.

\section{Simulations} \label{sec:sims}
In this section we describe the simulations used in this study.
We start by describing the different sets of initial conditions
(Gaussian, fixed and paired-fixed); we will then discuss the hydrodynamical
simulations and the method to simulate \lya\ forest sightlines from the simulations.

\subsection{Gaussian, Fixed and Paired-Fixed Random Fields}

In this paper we will follow the naming convention in \cite{Villaescusa2018},
which contains more detailed definitions including
power spectrum conventions.
Here we present a brief summary of the required definitions to interpret our results.

Given a density field $\rho(\vec{x})$, the density contrast is defined as
\be
 \delta(\vec{x})=\frac{\rho(\vec{x})-\bar{\rho}}{\bar{\rho}}~,
\ee
where $\bar{\rho}=\langle \rho(\vec{x})\rangle$.
In a simulation we take discretized values of these quantities at
the center of a cell $\vec{x_i}$, $\delta_i = \delta(\vec{x_i})$, and
define discrete Fourier modes $\tilde\delta_n$ as
\begin{equation}
 \tilde \delta_n = \sum_i \delta_i ~ e^{i \vec{x}_i \cdot \vec{k}_n}
    = A_n ~ e^{i \theta_n} ~,
\end{equation}
where $\vec{k}_n$ identifies the wavenumber, $A_n$ is the amplitude of the
discrete Fourier mode and $\theta_n$ is the phase.

For a Gaussian density field, each $\theta_n$ is a random variable distributed
uniformly between 0 and $2\pi$ and each $A_n$ follows a Rayleigh distribution
\be
p(A_n)~{\rm d}A_n=\frac{A_n}{\sigma_n^2}~e^{-A_n^2/2\sigma_n^2}{\rm d}A_n ~,
\label{Eq:Rayleigh}
\ee
with $\langle A_n \rangle = \sqrt{\frac{\pi}{2}} ~\sigma_n$, and $\sigma_n^2$
is proportional to the power spectrum $P(|\vec{k}_n|)$ 
(with the constant of proportionality
depending on the particular conventions adopted; see \cite{Villaescusa2018}).

For generating initial conditions (ICs) representing a discretized Gaussian density field for a simulation of the Universe, the mode amplitudes ($A_n$) and phases ($\theta_n$) are chosen randomly from their distributions above.
This randomness generates a source of variance in the final mock power spectrum from the simulation; the amplitudes $A_n$ propagate
directly, while the phases $\theta_n$ become significant during non-linear evolution because their values determine how
modes of different scales interact with each other.

In this paper, we explore alternative approaches to generating ICs such that these variance effects are minimised.
Similar to \citeauthor{Villaescusa2018}, we define Gaussian, paired Gaussian, fixed
and paired-fixed fields as follows:
\begin{itemize}
\item {\bf Gaussian field}: A field with $\tilde\delta_n=A_n~e^{i\theta_n}$,
where $A_n$ follows the Rayleigh distribution of Eq. \ref{Eq:Rayleigh} and $\theta_n$ is a random variable distributed uniformly between 0 and $2\pi$.
\item {\bf Paired Gaussian field}: A pair of Gaussian fields $\delta$ and $\delta'$,
where $\tilde\delta_n=A_n~e^{i\theta_n}$ and
$\tilde\delta_n'=A_n~e^{i(\theta_n+\pi)}=-\tilde\delta_n$;
the values of $A_n$ and $\theta_n$ are the same for the two fields
and $A_n$ is drawn from the Rayleigh distribution of Eq. \ref{Eq:Rayleigh}.
Because the leading-order uncertainties are typically from the amplitudes rather than
the phases, we do not use paired fields without fixing the amplitudes in this work.
\item {\bf Fixed field}: A field with $\tilde\delta_n=A_n~e^{i\theta_n}$,
where we fix $A_n = \sqrt{2}~\sigma_n$. The phase $\theta_n$ remains a uniform random
variable between $0$ and $2\pi$.
\item {\bf Paired-fixed field}: A pair of fields,
$\tilde\delta_n=A_n~e^{i\theta_n}$ and
$\tilde\delta_n'=A_n~e^{i(\theta_n+\pi)}=-\tilde\delta_n$,
where the values of $A_n$ are the same for the two fields
and fixed to $A_n = \sqrt{2}~\sigma_n$, and the values of $\theta_n$ are also the same for the two fields, as defined, rendering $\tilde\delta_n$ and $\tilde\delta_n'$ exactly out of phase with respect to each other. Here $\theta_n$ remains a uniform random variable between $0$ and $2\pi$.
\end{itemize}

The purpose of this paper is to compare the statistical accuracy of paired-fixed simulations with standard simulations (based on Gaussian fields), in the context of \lya\ forest power spectrum analyses.

\subsection{Hydrodynamical simulations} \label{sec:hydro}

We perform hydrodynamical simulations with the
\textsc{Gadget-3} code, a modified version of the publicly available code \textsc{Gadget-2} \citep{Springel2005}. The code incorporates radiative cooling by primordial hydrogen and helium, alongside heating by the UV background following the model of \cite{Viel_2013}. The \texttt{quick-lya} flag is enabled, so gas particles with overdensities larger than 1000 and temperatures below $10^5$ K are converted into collisionless particles \citep{Viel_2004}. This approach focuses the computational effort on low-density gas associated with the \lya\ forest, and is much more efficient than either allowing dense gas to accumulate or adopting a realistic galaxy formation and feedback model, to which the forest is largely insensitive. Thus, this computational efficiency has no significant effect on the accuracy of estimated \lya\ forest observables.

The initial conditions are generated at $z=99$ by displacing cold dark matter (CDM) and gas particles from their initial positions in a regular grid, according to the Zel'dovich approximation. We account for the different power spectra and growth factors of CDM and baryons by rescaling the $z=0$ linear results according to the procedure described in \cite{Zennaro_2016}.

Our simulations evolve $512^3$ CDM and $512^3$ gas particles from $z=99$ down to $z=2$ within a periodic box of $40 \hMpc$, storing snapshots at redshifts 4, 3 and 2. We run a total of 100 simulations (25 fixed pairs, and 50 standard simulations). In order to study the dependence of our results on volume we also run an identical set of simulations as those described above but with $256^3$ CDM and $256^3$ gas particles in a box of size $20\hMpc$.
        
The values of the cosmological parameters, the same for all simulations, are in good agreement with results from \cite{Planck}: $\Omega_{\rm m}=0.3175$, $\Omega_{\rm b}=0.049$, $\Omega_\nu=0$, $\Omega_\Lambda=0.6825$, $h=0.67$, $n_s=0.9624$, $\sigma_8=0.834$.

\subsection{Modeling \lya\ absorption} \label{sec:skewers}

In order to study {\lya} forest flux statistics we generate a set of
mock spectra containing only the {\lya} absorption line from neutral
hydrogen for each snapshot in our simulation suite. The spectra are
calculated on a square grid of 160 000 spectra for the 40 Mpc
box. Each spectrum extends the full length of the box,  and maintains the periodic
boundary conditions of the underlying simulation, with sizes in velocity space of \([3\,793, 4\,267,
4\,698]\,\mathrm{km}\,\mathrm{s}^{-1}\) respectively at \(z = [2, 3,
4]\). In each spectrum, the optical depth \(\tau\) is calculated by
taking the product of the column density of neutral hydrogen and the
atomic absorption coefficient for the \lya\ line \citep{humlicek79}
appropriately redshifted for both its cosmological and peculiar
velocities. Measurements are made in bins of velocity width \(10\,\mathrm{km}\,\mathrm{s}^{-1}\) using the code \texttt{fake\_spectra} \citep{bird17}.

Once we have computed the \lya\ optical depth, we derive the transmitted flux fraction (or flux for short):
\be
 F = e^{-\tau} ~.
\ee
We then calculate $\bar{F}$, the mean flux over all pixels in our box. Next, we adopt the common
practice of rescaling the optical depth of each simulation, $\tau \to \alpha \tau$ where $\alpha$
is chosen to fix $\bar{F}$ to a target value. This is normally adopted when making comparisons to data because it
reduces the impact of uncertainties in the UV background.
In our case, we set $\alpha$ in each simulation
such that $\bar{F}$ is fixed to the mean over all simulations. We verified that, if we do not
renormalize $\tau$, the benefits of paired and fixed simulations remain (in fact they even
increase further on small scales).
After renormalization, we calculate the fluctuations around the mean, $\delta_F$:
\be
 \delta_F = \frac{F}{\bar F} - 1~.
\ee

Finally, we calculate flux power spectra using the code \texttt{lyman\_alpha} \citep{Rogers2018a, Rogers2018b}. Studies of the \lya\ forest either adopt 1D or 3D power spectra depending on the density of available observational skewers; the former approach is computationally simpler and captures small-scale line-of-sight information, while the latter approach is required to capture larger-scale correlations that do not fit within a single quasar spectrum. In this work we will study the effects of paired-fixed simulations in both settings.

To create a mock power spectrum, the first step is to generate the Fourier transform (\(\tilde{\delta}_F\)) of the  fluctuation field \(\delta_F\). For 1D power spectra, the transformation is taken separately for each skewer along the line-of-sight direction; otherwise we apply the 3D transformation to the entire box.  In both cases we are able to exploit fast Fourier transforms since all velocity bins are of equal width and spectra are evenly sampled in the transverse direction. We then estimate power spectra by averaging $|\tilde\delta_F|^2$ over all skewers for the 1D power spectrum, or in $|\vec{k}|$ shells for the 3D power spectrum. 

When working with paired simulations, we average the power spectra to form a single estimate from each pair. This leaves us with $25$ independent estimates from our ensemble of $50$ paired-fixed simulations, a point that will be discussed further below.

\section{Results} \label{sec:results}
In this Section we will present the statistical performance of our different simulations in predicting \lya\ power spectra.
To quantify the performance of any given ensemble, we calculate the mean of the measured power across all realizations, $\bar{P}(k)$
\begin{equation}
\bar{P}(k) = \frac{1}{N_{\mathrm{est}}} \sum_{i=1}^{N_{\mathrm{est}}} P_i(k) \,,
\end{equation}
where $N_{\mathrm{est}}$ is the number of independent estimates of $\bar{P}(k)$ and $P_i(k)$ are the individual
estimates. We also calculate the single-estimate variance in $P(k)$:
\begin{equation}
\sigma^2(P(k)) = \frac{1}{N_{\mathrm{est}}-1} \sum_{i=1}^{N_{\mathrm{est}}} \left( P_i(k) - \bar{P}(k) \right)^2\,,
\end{equation}
As noted above, in the case of paired (or paired-fixed) simulations $N_{\mathrm{est}} = N_{\mathrm{sim}}/2$, while in the case of
a standard ensemble, $N_{\mathrm{est}} = N_{\mathrm{sim}}$ where in both cases $N_{\mathrm{sim}} = 50$, the number of simulations. This difference arises because paired power spectrum estimates are not independent and instead the {\it average} of each pair is regarded as providing a single sample $P_i(k)$.

We are actually interested in the expected uncertainty on the
mean power spectrum, $(\Delta \bar{P})^2$, estimated as
\begin{equation}
(\Delta \bar{P})^2 \equiv \frac{\sigma^2(P(k))}{N_{\mathrm{est}}}\textrm{.}\label{eq:Pbar-errors}
\end{equation}

The ratio of the standard to the paired-fixed uncertainties gives the most useful performance metric; in our case
where $N_{\mathrm{sim}}$ is the same between the two ensembles, this reduces to
\begin{equation}
\frac{(\Delta \bar{P}_{\mathrm{S}})^2}{(\Delta \bar{P}_{\mathrm{PF}})^2} = \frac{\sigma^2_\mathrm{S}(P(k))}{2\sigma^2_{\mathrm{PF}}(P(k))}\,,
\label{eq:varratio}
\end{equation}

where S and PF subscripts refer to the standard (Gaussian) and paired-fixed ensembles respectively. This ratio can also be thought of as the computational speed-up for achieving a fixed target accuracy (assuming the uncertainty on the mean power spectrum scales proportional to $\sqrt{N_{\mathrm{sim}}}$). It limits to unity when the paired-fixed approach is not providing any improvement.

From our paired-fixed simulations we can also form a fixed ensemble simply by
discarding one of each pair. Thus we will also show results for a fixed ensemble
with $N_{\mathrm{est}}=N_{\mathrm{sim}}=25$. Once again the estimated accuracy of
the mean power is given by Equation~(\ref{eq:varratio}), including the factor $2$
renormalisation for the relative number of estimates.

To ensure that the manipulations of the initial conditions do not introduce any bias,
we also calculate $\bar{P}_\mathrm{S}
- \bar{P}_\mathrm{PF}$. For finite $N_{\mathrm{sim}}$ this difference
contains a residual statistical uncertainty; one should expect
\begin{equation}
\langle (\bar{P}_\mathrm{S} - \bar{P}_\mathrm{PF})^2 \rangle = (\Delta
\bar P_{\mathrm{S}})^2 + (\Delta \bar P_{\mathrm{PF}})^2\label{eq:pbar-difference}\,,
\end{equation}
given the independence of the two ensembles. Therefore we check that $\bar{P}_\mathrm{S} - \bar{P}_\mathrm{PF}$ remains comparable in magnitude to this expected residual.

\subsection{Matter Power Spectrum}\label{subsec:Pm}

We start by discussing the matter power spectrum (i.e. the 3D power spectrum of the overdensity field without any transformation to \lya\ flux); \cite{Villaescusa2018} also discussed this quantity, but not at the redshifts in which we are interested in this work. Our simulations also have different baryonic physics implementations.

\begin{figure}
\centering
  \includegraphics[width=0.8 \textwidth]{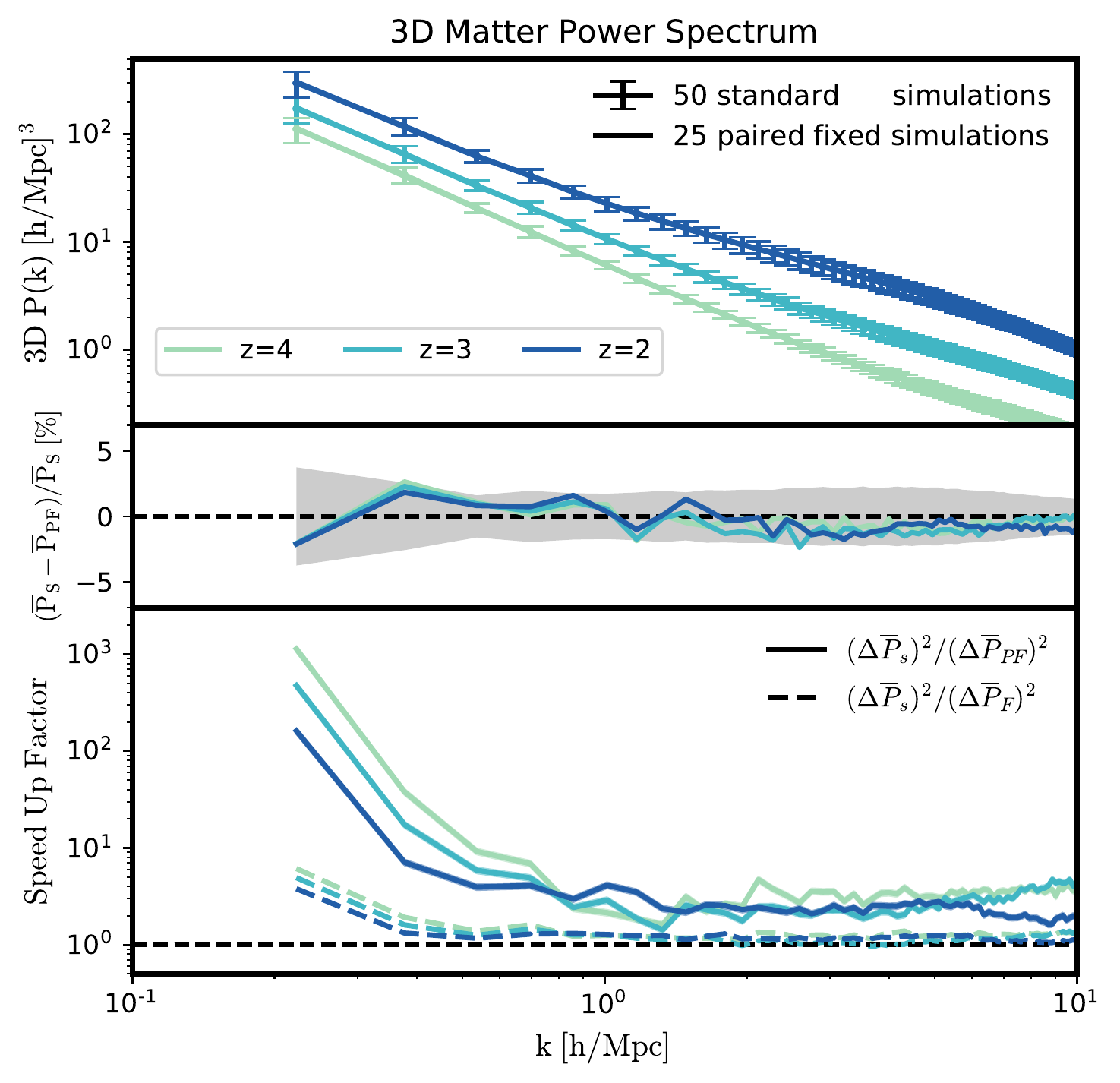}
\caption{Matter power spectrum measured at $z=2, 3$ and $4$.
The top panel shows the mean power spectrum in the two sets of simulations, the standard simulations, $\bar{P}_{\rm S}$, with the $1\sigma$ uncertainty on the mean, and the paired-fixed simulations, $\bar{P}_{\rm PF}$; the means are indistinguishable.
The middle panel shows the fractional difference between the mean power measured in the two sets of simulations, with a shaded region showing the expected $1\sigma$ uncertainty on that difference which is given by Eq. 10.
Across all scales, there is no evidence for bias in the estimate of the paired-fixed simulations relative to the traditional simulations.
The bottom panel shows the ratio between the uncertainty on the mean of the standard and paired-fixed simulations,
$(\Delta \bar P_\mathrm{S})^2 / (\Delta \bar P_\mathrm{PF})^2$, with dashed lines showing instead the effect of
fixing without pairs $(\Delta \bar P_\mathrm{S})^2 / (\Delta \bar P_\mathrm{F})^2$. These ratios summarise how fast the paired-fixed (or fixed) simulations converge on the mean power spectrum relative to the standard simulations.
As a typical example of improvement, for wavenumbers $k=0.25\ihMpc$ and $2\ihMpc$ at $z=3$, we find estimates of the matter power spectra converge $390$ and $2.0$ times faster in the paired-fixed simulations compared with the standard simulations.
}
\label{fig:pkm3d40}
\end{figure}

In Figure \ref{fig:pkm3d40} we show the matter power spectrum measured
at $z=2, 3, \mathrm{and} \;4$. In the top panel we show the average power spectrum
in the two ensembles: the set of 50 standard simulations
$\bar{P}_{\rm S}$ and the set of 25 paired-fixed simulations
$\bar{P}_{\rm PF}$.  Error bars show the uncertainties
$\Delta \bar P_{\rm S}$ in the mean of the standard simulations,
computed using Equation (\ref{eq:Pbar-errors}). The uncertainties are
small due to the large number of simulations. The uncertainties in the paired-fixed simulation are too small to be plotted in the top panel, as we will shortly discuss.

The middle panel shows the difference between the mean power
measured in the two sets of simulations, as a fraction of the total
power in the standard simulations. Additionally we show, as a grey band, the $\pm 1\sigma$
uncertainty at $z=3$\footnote{We find that our calculations of this expected
  scatter as a fraction of the mean power is almost independent of
  redshift, as the fractional uncertainties depend primarily just on the number
  of modes in each power spectrum bin.}  defined by
Equation~(\ref{eq:pbar-difference}). The two approaches to estimating
the non-linear power spectrum thus agree well, consistent with
\cite{Angulo2016} and \cite{Villaescusa2018}. In the bottom panel we quantify the statistical improvement achieved
by the paired-fixed simulations with respect to the standard
simulations.  We show the ratios of the uncertainty  on the means, comparing the standard
and paired-fixed simulations, $(\Delta \bar{P}_{\rm S})^2/(\Delta \bar
P_{\rm PF})^2$,
Equation (\ref{eq:varratio})\footnote{Note, this is the square of the quantity shown in the bottom panels in \cite{Villaescusa2018}.}. This ratio represents how quickly the
paired-fixed simulations converge on the true mean relative
to the standard simulations.  We also show the ratio of the uncertainty on the means of the standard simulations and fixed (but not paired)
simulations, shown as the dashed lines.  The dotted horizontal line
shows a value of 1, indicating the level where fixed and paired-fixed
simulations do not bring any statistical improvement over standard
simulations.
Comparing the solid and dashed lines shows that both fixing and
pairing the simulations contributes to reducing uncertainty on the mean power spectrum at low $k$.
As a typical example of improvement, for wavenumbers
$k=0.25\ihMpc$ and $2\ihMpc$ at $z=3$, we find
estimates of the matter power spectra converge $390$ and $2$ times
faster in paired-fixed simulations compared with standard
simulations.

Broadly this agrees with results from
\cite{Villaescusa2018} but, unlike in the earlier work, we find that
the factor $\simeq 2$ improvement in a paired-fixed simulation
continues to $k = 10\ihMpc$.  To isolate the cause of
this difference, we first considered the effect of box size.  Our
simulations are of boxes with side length $40\ihMpc$,
which is intermediate between boxes considered by
\cite{Villaescusa2018}. We therefore ran a suite of smaller
$20\ihMpc$ simulations for direct comparison to the smallest boxes of
that earlier work. However we found similar results at large $k$ in this additional
suite, and conclude that box size is not the driver for the different behavior. 
The only remaining difference between our present simulations and the hydrodynamic simulations of \cite{Villaescusa2018}
lies in the  drastically different baryonic physics implementation.
As discussed in Section
\ref{sec:hydro}, we do not implement feedback but instead use an
efficient prescription to convert high density gas into collisionless
particles. By contrast, \cite{Villaescusa2018} adopts a
state-of-the-art star formation and feedback prescription
which results in significant mass transport due to galactic
outflows. We verified that the $20\hMpc$ dark-matter-only runs of \cite{Villaescusa2018}, like our \lya\ runs,
also generate substantial improvements from pairing and fixing at high $k$.
It therefore seems that potential efficiency gains at $k \gg 1 \ihMpc$
can be erased by stochastic noise from sufficiently
energetic feedback and winds.

Our practical conclusions will be insensitive to this regime.
Uncertainties
at high $k$ are dominated by observational rather than theoretical
uncertainty, due to the rapidly shrinking absolute magnitude of
simulation sample variance. Because the sample variance is intrinsically small, it is only important that
the power spectrum remains unbiased at these scales.

\subsection{3D \lya\ Power Spectrum}\label{subsec:P3D}

We next turn our attention to the 3D \lya\ power spectrum.  On large,
quasi-linear scales, the 3D \lya\ power spectrum is proportional to the matter power
spectrum, with an amplitude set by the scale-independent bias
parameter, and an angular dependence described by the \cite{Kaiser1987}
model of linear redshift-space distortions. In this work,
we include these distortions but consider the angle-averaged (i.e. monopole)
power spectrum rather than divide the power spectrum into angular bins. 

\begin{figure}[t]
\centering
  \includegraphics[width=0.8\textwidth]{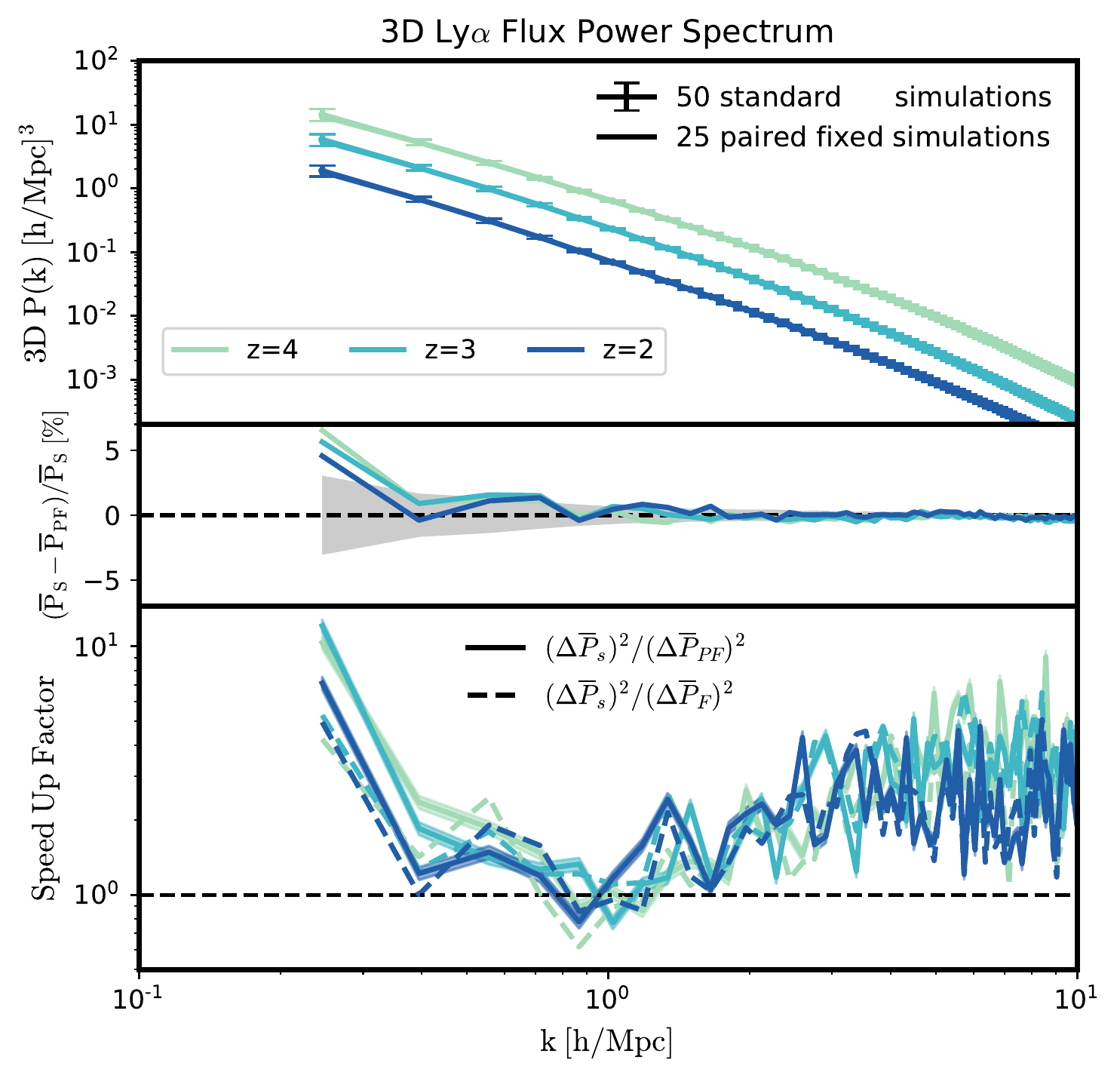}
\caption{Same as Figure \ref{fig:pkm3d40}, but for the 3D \lya\ flux power spectrum. As a typical example of improvement, for wavenumbers $k=0.25\ihMpc$ and $2\ihMpc$ at $z=3$, we find estimates of the 3D \lya\ flux power spectra converge $12$ and $2.0$ times faster in a paired-fixed ensemble compared with a standard ensemble.
 }
\label{fig:varratio3d40}
\end{figure}

We present the measured 3D \lya\ forest power spectrum in Figure
\ref{fig:varratio3d40}.  The panels are computed and arranged as in
Figure \ref{fig:pkm3d40}, but starting from the flux $\delta_F$ instead of
the overdensity $\delta$.
  In the top panel the
uncertainties on the mean power spectrum of the standard simulations
are again small due to the large number of simulations. The middle panel shows the fractional difference between the
paired-fixed and standard simulation power spectra, with the grey
shaded region again indicating the expected $1\sigma$ scatter.
Several measured points scatter outside this envelope (most notably at
$k<0.3\ihMpc$) but this is consistent with a statistical
fluctuation driven by scatter in the standard ensemble. At high $k$
the difference in the mean is sub-percent. Comparing the central panels of Figures~\ref{fig:pkm3d40} and~\ref{fig:varratio3d40} reveals that the shape of the $1\sigma$ contour is different between the \lya\ and matter power cases. At high $k$, the matter power enters a highly non-linear regime where modes are strongly coupled, whereas
the flux arises from low density clouds that are still quasi-linear. In the case of flux, modes are near-decoupled and the
uncertainty decays proportionally to the increasing number of modes per $k$ shell.

The bottom panel of Figure~\ref{fig:varratio3d40} quantifies the
statistical improvement achieved by the paired-fixed
simulations with respect to the standard simulations, shown as the solid line. The
dashed line represents simulations that are just fixed.  The dotted
horizontal line shows a value of 1, indicating the level where fixed
and paired-fixed simulations do not bring any statistical improvement
over standard simulations. As a typical example of improvement, for
wavenumbers $k=0.25\ihMpc$ and $2\ihMpc$ at $z=3$,
we find estimates of the 3D \lya\ flux power spectra converge $12$
and $2$ times faster in a paired-fixed ensemble compared with a
standard ensemble.

There is a minimum level of improvement at
$k \simeq 1\ihMpc$ where the paired-fixed approach does not outperform the standard ensemble at $z=3$. The relative performance
improvements then {\it increases} with increasing $k$ beyond this
point, which is surprising. Previously it has been argued that the
improvements generated by the paired-fixed approach can be understood
within the framework of standard perturbation theory
\citep{Angulo2016,Pontzen2016} which more obviously applies to
intermediate quasi-linear scales than the $k>1\ihMpc$
regime.

At present we cannot fully explain why pairing and fixing improves the
\lya\ power spectrum accuracy in this regime. It might be that
super-sample covariance terms couple high $k$ to low
$k$ in a unique way given that the forest is dominated by underdense
regions (unlike the matter power spectrum; \citealp{Takada2013}). 
However, we did not
investigate further for this work because the improvements are not
particularly relevant for observational studies; the absolute
magnitude of the model uncertainties in the $k>1\ihMpc$ rapidly become
smaller than observational uncertainties (see top panel of Figure
\ref{fig:varratio3d40} for a visualization of the magnitude of the model uncertainties). Thus the improvements in accuracy at low $k$
are of most practical benefit to future work.

\subsection{1D \lya\ Power Spectrum}\label{subsec:P1D}

Most hydrodynamical simulations of the \lya\ forest have been used to
predict the 1D power spectrum
\citep{McDonald2005a,Viel2006a,Borde2014,Lukic2015,Walther2018b}.
As described in Section \ref{sec:skewers}, the
1D spectrum captures small-scale information along the line of sight
without the computational complexity of cross-correlating between
different skewers.

\begin{figure}
\centering
  \includegraphics[width=0.8\textwidth]{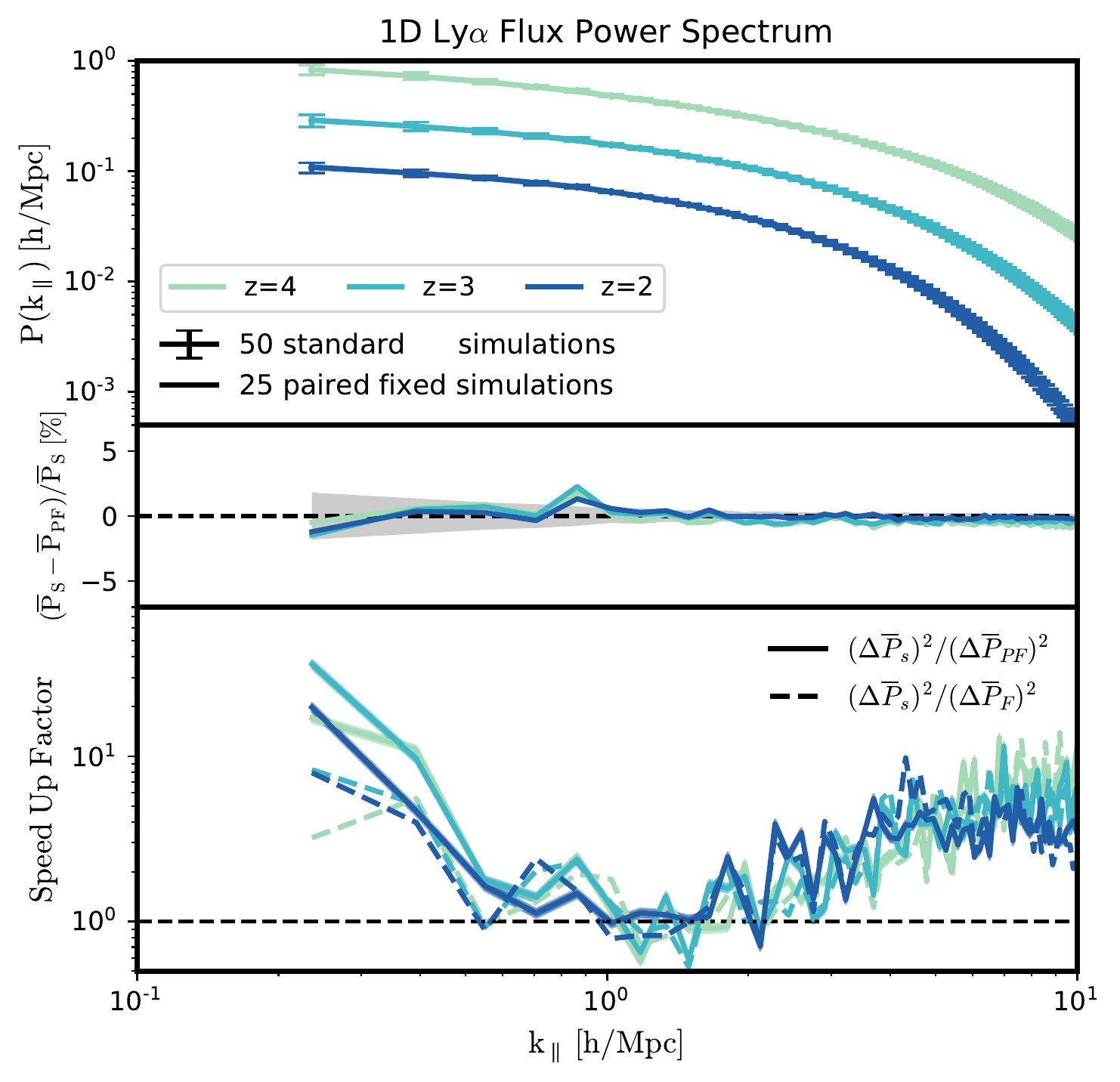}
  \caption{Same as Figure \ref{fig:pkm3d40}, but for the 1D \lya\ flux power spectrum.
   The line-of-sight wavenumber $\kpar$ corresponds to an integral over many
   3D wavenumbers $k$, so there is not a direct correspondence between the scales. As a typical example of improvement, for wavenumbers $k_{\parallel}=0.25\ihMpc$ and $2\ihMpc$ at $z=3$, we find estimates of the 1D \lya\ flux power spectra converge $34$ and $1.7$ times faster in a paired-fixed ensemble compared with a standard ensemble.
}
\label{fig:varratio1d40}
\end{figure}

We present the measured 1D \lya\ power spectrum in Figure
\ref{fig:varratio1d40}, using again the same panel layout as Figure
\ref{fig:pkm3d40}.  In the top panel the uncertainties on the mean
power spectrum of the standard simulations are, as usual, small due to the
large number of simulations. The middle panel shows the fractional
uncertainty in the paired-fixed simulations relative to the
standard simulations, with the grey shaded region indicating the $1\sigma$
uncertainty on this difference.
Once again there is overall good agreement, with the strongest
deviation arising at $k=0.9\ihMpc$ at a significance of
almost $2\sigma$. However the difference between our standard and
paired-fixed ensemble remains less than $2\%$, and (given the large number
of independent $k$ bins) we believe the
difference to be consistent with the expected level of statistical
fluctuations.

The bottom panel quantifies the statistical improvement achieved by
the paired-fixed simulations (solid line) with respect to the standard
simulations. The dashed line represents simulations that are fixed but
not paired. As a typical example of improvement, for wavenumbers
$k_{\parallel}=0.25\ihMpc$ and $2\ihMpc$ at $z=3$,
we find estimates of the 1D \lya\ flux power spectra converge $34$ and
$1.7$ times faster in a paired-fixed ensemble compared with a
standard ensemble.
As with the 3D \lya\ flux power (Section
\ref{subsec:P3D}) there is a steadily increasing improvement at high
$k$, forming a local minimum at $k_{\parallel} \simeq 1\ihMpc$;
since the 1D power spectrum mixes multiple modes from the 3D power
spectrum, this is consistent with projecting the improvement discussed
in Section \ref{subsec:P3D}.

\section{Discussion and Conclusions} \label{sec:conc}

We have shown that using the method of paired-fixed simulations reduces the uncertainty of the mean mock \lya\ forest power spectrum measured in hydrodynamical simulations, and therefore requires less computing time to estimate the expected value of the considered quantity, which is needed to evaluate the likelihood of the data.
As a typical example of improvement, for wavenumbers $k=0.25\ihMpc$ at $z=3$, we find estimates of the 1D and 3D power spectra converge $34$ and $12$ times faster in a paired-fixed ensemble compared with a standard ensemble.
The largest improvements are at small k, where model uncertainties are larger than observational uncertainties. The improvements are minimal at $k=1\ihMpc$, but at these scales the observational uncertainties are larger than the model uncertainties, so reduced uncertainty on the mean mock power spectrum is unnecessary. This suggests that running a large simulation box size with paired-fixed ICs will provide the most accurate mock \lya\ forest power spectrum over the largest range of scales.

By reducing the computational time required to achieve a target accuracy for mock power spectra, the method frees up resources for a more thorough exploration of astrophysical and cosmological parameters. It is essential to be able to sample efficiently over this space, with at least three parameters describing the shape and amplitude of the input linear power spectrum and at least two for astrophysical effects to span the reionisation redshift and level of heat injection.
In future, likelihoods for datasets such as eBOSS and DESI will likely take advantage of emulators which
interpolate predicted power spectra within these high-dimensional parameter spaces (\cite{Heitmann16,Walther2018b}, Rogers et al in prep, Bird et al in prep). In this context, freeing up CPU time by using the paired-fixed approach will allow for a denser sampling of training points for the emulator. That in turn will beat down interpolation errors and thus lead to more accurate inferences from forthcoming data.

\section*{Acknowledgements} \label{sec:ack}

It is a pleasure to thank An\v{z}e Slosar for useful discussions. The work of LA, FVN and SG is supported by the Simons Foundation. AP was supported by the Royal Society. AFR was supported by an STFC Ernest Rutherford Fellowship, grant reference ST/N003853/1. AP and AFR were further supported by STFC Consolidated Grant no ST/R000476/1.
This work was partially enabled by funding from the University College London (UCL) Cosmoparticle Initiative.
This work was supported by collaborative
visits funded by the Cosmology and Astroparticle Student and Postdoc Exchange
Network (CASPEN).
The simulations were run on the Gordon cluster at the San Diego Supercomputer Center supported by the Simons Foundation.
\software{
The code used in this project is available from
\url{https://github.com/andersdot/LyA-InvertPhase}.
This research utilized the following open-source \textsl{Python} packages:
    \textsl{Astropy} (\citealt{Astropy-Collaboration:2013}),
    \textsl{scipy}(\citealt{scipy:2001}),
    \textsl{matplotlib} (\citealt{Hunter:2007}),
    \textsl{fake\_spectra} (\citealt{bird17}) and
    \textsl{numpy} (\citealt{Van-der-Walt:2011}).
It also utilized the python pylians libraries, publicly available at \url{https://github.com/ franciscovillaescusa/Pylians}, and \textsl{lyman\_alpha} \citep{Rogers2018a, Rogers2018b}}

\bibliography{reference}

\end{document}